\def\be{\begin{equation}}
\def\ee{\end{equation}}
\def\bea{\begin{eqnarray}}
\def\eea{\end{eqnarray}}
\def\bse{\begin{subequations}}
\def\ese{\end{subequations}}

\documentclass[prb,twocolumn,showpacs,preprintnumbers,amsmath,amssymb]{revtex4}
\usepackage{graphicx}
\usepackage{dcolumn}
\usepackage{bm}
\usepackage{wasysym}

\begin{document}
\title{Emulating Non-Abelian Topological Matter in Cold Atom Optical Lattices
}
\author{V. W. Scarola and S. Das Sarma }
\affiliation{Condensed Matter Theory Center, Department of
Physics, University of Maryland, College Park, MD 20742}

\date{\today}
\begin{abstract}
Certain proposed extended Bose-Hubbard models may exhibit topologically 
ordered ground states with excitations obeying non-Abelian 
braid statistics.  A sufficient tuning of Hubbard parameters could yield  
excitation braiding rules allowing implementation of a universal set of topologically protected 
quantum gates.  We discuss potential difficulties in realizing a model with a 
proposed non-Abelian topologically ordered 
ground state using optical lattices containing bosonic dipoles.  Our direct implementation scheme does not 
realize the necessary anisotropic hopping, anisotropic interactions, and low temperatures.
\end{abstract}

\pacs{03.75.Lm}

\maketitle

\section{\label{Introduction} Introduction}

Topological matter is operationally defined \cite{RMP} as a 
two-dimensional quantum many-body system with a non-trivial ground 
state degeneracy immune to weak local perturbations.  
The existence of an excitation gap separating  
the ground state from the low-lying excitations guarantees  
quantum immunity of the ground state degeneracy against local perturbations.  
Topological matter sustains, in general, two types of excitations.  
With type I (type II) topological order quasiparticles obey Abelian (non-Abelian) 
statistics.  Physically braiding excitations in type I and type II topological matter 
modifies the many-body wavefunction by a phase and non-trivial matrix, respectively.  Braiding 
non-Abelian quasiparticles may enable fault-tolerant topological quantum computation \cite{Kitaev,RMP}, 
reducing stringent quantum error correction procedures required of ordinary 
qubit-based quantum computation.  We further  
classify type II topological order based on potential application in topological 
quantum computation.  Braiding 
a small number of excitations in type IIa topologically ordered matter does not yield a universal set of 
quantum gates necessary for implementing all quantum codes.  Braiding in 
type IIa matter must be supplemented by unprotected quantum gates in order to implement 
an arbitrary quantum code thereby offering a ``partial'' topological immunity to 
weak local noise.  Braiding excitations in type IIb systems, in contrast, yields a universal 
set of quantum gates offering ``full'' topological immunity in implementing an 
arbitrary quantum code with braid operations.  Models demonstrating type IIb 
topological order incorporate additional 
complexity to accommodate a universal set of gates in the excitation braid structure.  

Although a great deal is known theoretically 
about topological matter in the effective field theory sense, the necessary 
conditions for the emergence of topological order and non-Abelian 
quasiparticle statistics in real materials are not entirely known.  A current 
candidate of type IIa topological order is the so-called $\nu=5/2$ fractional 
quantum Hall state occurring at $mK$ temperatures in GaAs based 
high mobility two-dimensional electron systems 
subjected to a strong external magnetic field.  However, no  
direct experimental evidence exists establishing the non-Abelian topological 
nature of the $\nu=5/2$ fractional quantum Hall state \cite{DasSarma} or any real material.  
Similarly, it is conjectured \cite{Read} that the fragile and rarely
observed 12/5 fractional quantum Hall state may be type IIb with anyonic
quasiparticles (the so-called "Fibbonacci anyons") suitable for universal
quantum computation, but little is known about the nature of this very weak
state.  It is, however, well-accepted \cite{Jain_book,Jain_anyon} that 
the quasiparticle excitations of the well-known 
fractional quantum Hall states (i.e. 1/3, 2/3, 2/5, 3/7, and
so on) are type I.

Given the absence of any experimentally definitive topological system, much theoretical 
work has gone into effective theoretical lattice models which have topological 
many-body ground states.  The most famous example of a type IIa topological lattice model in 
the quantum information context is Kitaev's toric code \cite{Kitaev}, a two-dimensional spin lattice model.  
Another such example, of interest to our work 
presented in this paper, is a Bose-Hubbard lattice model on a kagome lattice with 
extended and ring exchange interaction terms which
has been proposed as a model carrying type IIb topological order \cite{Freedman0,Freedman}.  One 
possible advantage of the topological lattice models is that, although 
these lattice models are highly contrived from the solid state physics perspective 
(and their applicability to real solid state materials is completely unknown), it is,  
in principle, possible to imagine emulating them on cold atom optical lattices similar to 
what has been already experimentally achieved in realizations of the on-site  
Bose-Hubbard model with bosonic optical lattices\cite{Jaksch,Greiner,Bloch_Review}.  
Motivated by the remarkable success of quantum analog simulation of 
solid state models \cite{Lewenstein} several recent theoretical proposals have been made to  
create \cite{Duan,Zoller_polar} and manipulate \cite{Zhang} 
cold atom optical lattices emulating Kitaev spin models\cite{Kitaev,Kitaev2}.  

In this paper 
we theoretically identify key issues in optical lattice emulation of the 
extended Bose-Hubbard model originally considered as a quasi-realistic lattice 
model by Freedman \textit{et al.} \cite{Freedman0,Freedman}.  While 
Refs.~\onlinecite{Freedman0} and ~\onlinecite{Freedman} 
study this model independent of specific experiments, it is natural to ask if experiments 
can approximate such a model.  We analyze a ``direct''  
optical lattice emulation of the topological extended Bose-Hubbard 
model with dipoles.  By ``direct'' we mean optical lattices formed from interfering 
standing wave lasers containing bosons with a dipolar interaction.  
Our direct scheme combines recent work including optical lattice experiments \cite{Porto}, 
proposals \cite{Santos}, and results showing Bose-Einstein condensation with 
dipolar atoms \cite{Chromium}.  We find that our direct implementation scheme 
would be extremely difficult, if not impossible, even as a matter 
of principle because the combined parameter constraints on the 
hopping, various interaction terms, superexchange, temperature, single band restriction, 
and chemical potential required to reach the proposed topological regime are, for all practical purposes, 
mutually exclusive.  Specifically, we find that: 
i)  Our lattice setup significantly alters chemical potentials to yield prohibitively 
long hopping times between sites (See Fig.~\ref{fig2}) 
ii) Band effects in our lattice setup, combined with the isotropic dipolar 
interaction, do not induce a sufficient 
anisotropy in the interaction (we require an order of magnitude relative 
anisotropy among next nearest lattice sites 
but we achieve 8$\%$ at best),  
iii) The realization of many-body superexchange 
in harmonic optical lattices presents stringent constraints on 
temperature, $T$ (See Fig.~\ref{fig3}).

The plan of the paper is as follows:  In section \ref{Model} we review 
relevant aspects of the extended Bose-Hubbard 
model of Refs.~\onlinecite{Freedman0} and \onlinecite{Freedman}.  In 
section \ref{Dipoles} we calculate the Hubbard parameters 
for a direct implementation of an extended Bose-Hubbard model with 
dipoles confined in a kagome optical lattice.  We attempt to modify the lattice to tune 
Hubbard parameters.  We find that our lattice ``coloring'' scheme does not yield an 
appropriate set of Hubbard parameters.  In section \ref{Exchange} we discuss practical issues in 
realizing low temperature superexchange with harmonic optical lattices.  We 
find that weak (harmonic) site confinement places prohibitive constraints on the temperature.  In 
section \ref{Conclusion} we summarize difficulties we encountered 
in implementing the topological extended Bose-Hubbard model.  We emphasize that 
our work here underscores difficulties with our direct implementation 
scheme using dipoles.   
  
\section{\label{Model} Model}

\begin{figure}
\includegraphics[width=3in]{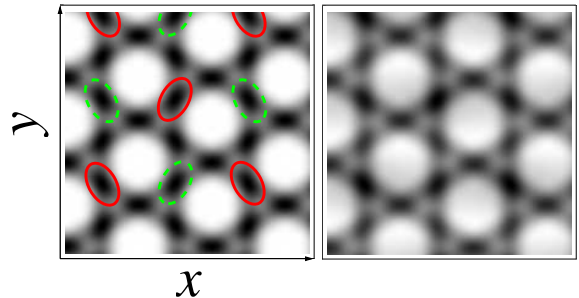}
\caption{Left panel:  Grey scale plot of the kagome lattice, defined by 
Eq.~\ref{Vkag}, in the $x-y$ plane.  Dark regions indicate sites.  Sites encircled with  
solid (dashed) lines indicate red, r, (green, g,) sites.  
Sites not encircled indicate black, b, sites. 
Right panel:  The same but with an additional potential (Eq. ~\ref{Vshift}) 
added to color the lattice with $I_s=0.3$.  
}
\label{fig1}
\end{figure}

We consider an optical lattice setup which 
approximates a single band Bose-Hubbard model.  In 
Refs.~\onlinecite{Freedman0} and \onlinecite{Freedman} it was argued that 
bosons hopping in a 2D kagome lattice (see left 
panel of Fig.~\ref{fig1}) can realize 
quantum topologically ordered ground states of type IIb, with Bose-Hubbard
parameters tuned to specific values.  
The proposal presents the following model with 
an additional ring exchange term\cite{Freedman0,Freedman}:
\begin{eqnarray}
H_{\text{TB}}&=&
-\sum_{i}\mu_i n_i
-\sum_{<i,j>}t_{i,j}\left(b_{i}^{\dagger}b_{j}^{\vphantom{\dagger}}+h.c.\right) 
\nonumber\\
&+&\frac{U_0}{2}\sum_{i}n_i(n_i-1)+U_{\hexagon}\sum_{(i,j)\in\hexagon}n_i n_j 
\nonumber\\
&+&\sum_{(i,j)\in\bowtie,\notin\hexagon}V_{i,j}n_i n_j.
\label{HamiltonianTB}
\end{eqnarray}
The $\mu_i$ define site dependent chemical potentials and $n_i$ is the number operator.  The second term 
indicates bond-specific nearest neighbor hoppings with energy gain $t_{ij}$, 
where $b_j$ annihilates a boson at the site $j$.  $U_0$ represents an onsite 
interaction energy penalty assumed to 
be the largest energy scale in the single band limit.  $U_{\hexagon}$ is an energy 
penalty between particles on a given hexagon 
and is tuned to be large thereby preventing double occupancy 
of hexagons.  With one particle per hexagon and uniform hopping one finds a  
degenerate manifold of boson configurations that can be thought of as 
dimer configurations where each dimer lies along a line connecting the center of 
each hexagon.  The degeneracy can be lifted by modifying hoppings 
or adding inter-hexagon interaction 
energy penalties (the last term in Eq.~\ref{HamiltonianTB}).  The latter  
impose an energy cost (to be matched with superexchange interaction energies) between 
next-nearest neighbors lying along 
bow-ties not within the same hexagon.  A proposed tuning 
of $\mu_i$, $t_{i,j}$, $V_{i,j}$, and an additional multi-site ring exchange 
term (not discussed here) drive the system towards type IIb topological order
\cite{Freedman0,Freedman}.  In the same context a simpler set of conditions 
were proposed for realizing type I ground states (the $k=1$ topological phase 
realized with Eqs. 9-14 in Ref.~\onlinecite{Freedman}).
Some of these conditions are summarized below:
\begin{eqnarray}
\epsilon&=&t^{\text{b}}_{\text{gb}}/U_{\hexagon}=t^{\text{r}}_{\text{bb}}/U_{\hexagon}
\nonumber \\
&=&t^{\text{b}}_{\text{rb}}/(cU_{\hexagon})
\nonumber \\
V_{gb}^b/U_{\hexagon}&=&V_{bb}^b/U_{\hexagon}=2ac\epsilon^2 
\nonumber \\
V_{rb}^b/U_{\hexagon}&=&V_{rg}^b/U_{\hexagon}=2c\epsilon^2/a
\nonumber \\
V_{bb}^r/U_{\hexagon}&=&V_{bb}^g/U_{\hexagon}=2\epsilon^2
\nonumber \\
V_{i,j}&\ll& U_{\hexagon}
\nonumber  \\
t &\ll& U_{\hexagon}
\nonumber  \\
T&\ll& t^2/U_{\hexagon} 
\nonumber  \\
\text{Interactions}&\ll&\Delta.
\label{conditions}
\end{eqnarray}
Here $t^{\gamma}_{\alpha \beta}$ indicates hopping between 
sites colored $\alpha\in \{\text{b,g,r}\}$ and $\beta\in \{\text{b,g,r}\}$ with 
$\gamma\in\{\text{b,g,r}\}$  the color of the site opposite $\alpha$ and 
$\beta$ in the corresponding triangle, Fig~\ref{fig1}.  
$V^{\gamma}_{\alpha \beta}$ indicates $V_{ij}$ where
$\alpha$ and $\beta$ represent  
the colors of next-nearest neighbors $i$ and $j$ while $\gamma$ is the color of the 
site between them. 
$\epsilon$ is a small positive 
number and $c$ and $a$ are constants.  
A modification of these conditions, involving additional complexity, may, as proposed 
in Refs.~\onlinecite{Freedman0} and \onlinecite{Freedman}, drive the system towards type IIb 
topological order.  The last two conditions depend on temperature and the 
energy splitting between the lowest and first excited band, $\Delta$.  
They are practical constraints for a direct realization of a 
Hubbard model in a single band superexchange limit with low temperatures.  

\section{\label{Dipoles} An Optical Lattice of Dipoles}

We now consider key difficulties in directly implementing 
the above tight binding model using bosons confined to optical 
lattices.  Optical lattices offer a tunable environment 
free from defects, impurities, and lattice phonons.  Implementations 
of Bose-Hubbard models have realized low temperature ($T \lesssim t$) superfluid and Mott 
phases \cite{Jaksch,Greiner,Bloch_Review}.  These Bose-Hubbard systems 
have been realized with Alkali atoms parameterized by a zero-range 
contact interaction.  Effectively contributing only $U_0$ in 
Eq.~\ref{HamiltonianTB}.  Recent work seeks to extended the range of interaction between 
particles in optical lattices by promoting bosons to higher bands 
\cite{Scarola,Mueller}. 
Promoting bosons to higher bands uses band effects to expand the contact interaction 
to include a weak nearest neighbor interaction.  
In implementing 
Eq.~\ref{HamiltonianTB} we need interactions to be 
tuned over several nearest and next nearest neighbors.  Work in a different 
system\cite{Goral,Chromium} highlights the possibility of 
confining dipoles to optical lattices.  Dipoles in lattices generate 
nearest and next-nearest neighbor interaction terms in 
Hubbard models.  Magnetic dipoles (e.g. $^{52}$Cr) have a weak dipolar 
component (See, e.g., Ref.~\onlinecite{Menotti} for estimates).  For example, 
lattice depths yielding a hopping of $0.1 E_R$ leave a nearest neighbor interaction 
below $3\times 10^{-4} E_R$ for $t\approx U_0$ with $^{52}$Cr.  But recent 
proposals indicate that molecules with electric dipolar moments 
may yield stronger dipolar contributions to the 
interaction (See, e.g., Refs.~\onlinecite{Zollerdipole1,Santos}).     

To be specific
we assume that bosons with a strong dipolar interaction can
be confined to optical lattices.  The Hamiltonian for interacting particles of mass 
$m$ in a single particle potential defined by an optical lattice, $V_{\text{OL}}$, is given by:  
\begin{eqnarray}
H&=& \sum_k \left[-\frac{\hbar^2}{2m}\bm{\nabla}^2_{\bm{r}_k} 
+ V_{\text{OL}}(\bm{r}_k)\right] 
\nonumber \\
&+&\frac{1}{2}\sum_{k\neq l}V_{\text{Int}}(|\bm{r}_k-\bm{r}_l|).
\label{Hamiltonian}
\end{eqnarray}
To define a two-dimensional 
lattice in the $x-y$ plane we assume confinement along the 
$z$-direction, $V_z(z)$, sufficiently deep to prevent excitations 
out of the lowest confined state in the $z$ direction.  The $z$ 
component of the wavefunction can be approximated by a Gaussian:
$\phi(z)=(2/\pi l^2)^{1/4}\exp{(-z^2/l^2)},$
where $l$ is defined by the confinement frequency along the $z$ direction. 
The Fourier transform 
of the dipolar interaction (excluding the contact interaction),  
with the dipoles oriented along the $z$ direction,
is then\cite{Pedri}:
\begin{eqnarray}
\tilde{V}_{\text{Int}}^{2D}(\bm{k}_p)&=&g\left[\frac{1}{l\sqrt{\pi}}-\frac{3k_p}{2}\exp({l^2 k_p^2})
\text{Erfc}(l k_p)\right],
\end{eqnarray}
where $g$ is an interaction parameter, $\bm{k}_p=(k_x,k_y)$, and 
$\text{Erfc}$ is the complimentary error function.  In real space 
this interaction is isotropic, decays as $r^{-3}$ at large distances, and has 
its short range part suppressed 
by the $z$ extent of the wavefunction.  We use the
above interaction 
(including a contact interaction) to calculate the Hubbard parameters.  We assume 
that the strength of the contact interaction and the dipolar interaction, $g$, are 
independently tunable over an arbitrary range.

The kagome lattice can be defined using six counter-propagating laser 
beams of wavevector $k$.  By interfering the beams at 
specific angles the resulting potential is given by, 
$V_{\text{OL}}=I_{0}V_{\text{kag}}+V_z$, where
\cite{Santos}:
\begin{eqnarray}
V_{\text{kag}}(\bm{r})&=&
\sum_{\alpha=1-3}[\cos\left(\bm{k}_{\alpha}\cdot\bm{r}+\phi_{\alpha}/2\right)
\nonumber \\
&+&2\cos\left(\bm{k}_{\alpha}\cdot\bm{r}/3+\phi_{\alpha}/6\right)]^2.
\label{Vkag}
\end{eqnarray}
With 
$\bm{k}_1=k(-1/2,-\sqrt{3}/2),
\bm{k}_2=k(1/2,-\sqrt{3}/2),
\bm{k}_3=k(1,0),
\phi_1=-\phi_2=\phi_3=\pi$, and 
$I_0=-E_{\text{R}}/2$
the potential minima define an ``isotropic'' kagome 
lattice (see the left panel in Fig.~\ref{fig1}).  
We define $E_{\text{R}}=h^2/8ma^2$ with $a=\pi/k$.  We use the tight binding 
basis of localized states (Wannier functions) centered 
at sites arranged in a kagome lattice.  To calculate the Hubbard 
parameters we approximate the Wannier functions by Gaussians in a variational 
ansatz \cite{Santos}.  We minimize the single particle part of 
Eq.~\ref{Hamiltonian} with respect to four variational parameters, the Gaussian 
width and location, near each distinct site.  
For large lattice depths  
the Gaussian approximation provides a reasonable estimate 
of tight binding parameters calculated from a full band theory treatment 
\cite{Santos}.  The large number of interfering beams 
used to define the kagome lattice (Eq.~\ref{Vkag}) 
superpose to yield deep site confine for $ \vert I_0 \vert \sim  0.5 E_{\text{R}}$.  This is contrast 
to square optical lattice geometries \cite{Jaksch} that are safely in the single-band, tight-binding limit 
for much larger lattice depths (near $10 E_{\text{R}}$).  
 
For dipoles in an isotropic kagome 
lattice defined by $V_{\text{kag}}$ the hopping, chemical potential, and 
interactions at equal distances are all uniform throughout the lattice.  
Note that the spatial distance between site pairs across a hexagon and along a bow-tie are 
the same.  We find, by direct calculation, that the 
requirement $U_{\hexagon}\gg V_{i,j}$ is therefore not satisfied 
by a kagome lattice of dipoles 
with a 2D spatially isotropic interaction.  Furthermore the interaction between 
dipoles decays as $r^{-3}$ leaving corrections at large distances.  We 
ignore corrections to the interaction beyond the next-nearest neighbor.  In 
phases with an energy gap ($\sim \mathcal{O}(\epsilon^2)$ in our example here) 
we assume that the long range correction terms are much smaller than the gap thereby preventing 
a phase transition.  This approximation depends on a theoretical unknown,  
the stability of the gapped topological phase with respect to perturbations.  Emulating a topological 
phase in optical lattices would present us with an experimental tool to probe stability.  We therefore 
ask if some of the conditions imposed on $V_{i,j}$ (Eqs.~\ref{conditions}) can be 
partially met with the short range part of the dipolar interaction.  Here we must 
match superexchange terms $\mathcal{O}(t^2/U_{\hexagon})$  with interaction terms, $V_{i,j}$. 

\begin{figure}
\includegraphics[width=3in]{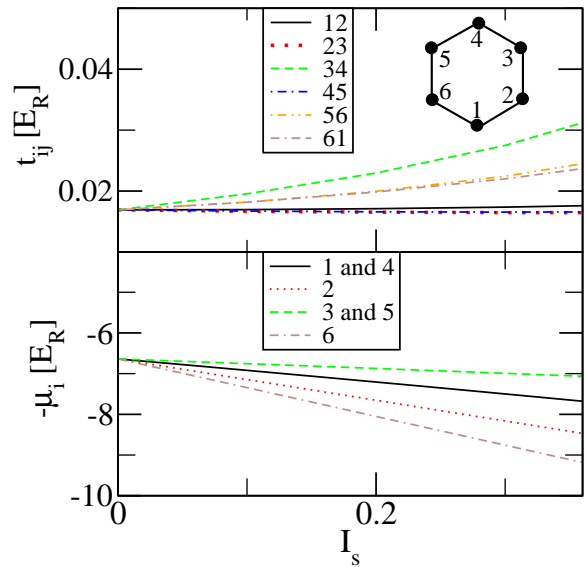}
\caption{Top panel:  Hopping between site pairs around a hexagon 
as a function of the shift  potential strength, $I_s$. 
The lower left hexagon from  Fig.~\ref{fig1} is labeled $1-6$, where $2$ is 
a green site.  $I_s$ approximately tunes $c$ in Eqs.~\ref{conditions}.  
Bottom panel:  Chemical potential around the hexagon resulting from the addition 
of the shift potential to the kagome.  
}
\label{fig2}
\end{figure}

We attempt to impose anisotropy in the lattice by tuning $c$ in 
Eqs.~\ref{conditions} with a shifted potential.  Applying an additional 
set of counter-propagating beams can, at least in principle, be used to 
``color'' the otherwise uniform kagome lattice.
Recent experiments have colored a two-dimensional 
square lattice \cite{Porto}.  We have here a 
dual goal: i) Can we color hoppings in accord with Eqs.~\ref{conditions}?
 ii) Do band effects in the colored lattice induce substantial 
(order of magnitude) anisotropies in the interaction?  We 
study an applied shift potential established by 
four additional beams to modify the hopping in an anisotropic fashion:    
\begin{eqnarray}
V_{\text{s}}(\bm{r})= 
[\cos\left(\bm{k}_{4}\cdot(\bm{r}-\bm{R}_1)\right)
+A\cos\left(\bm{k}_{5}\cdot(\bm{r}-\bm{R}_1)\right)]^2\nonumber \\
+[\cos\left(\bm{k}_{4}\cdot(\bm{r}-\bm{R}_2)\right)
+A'\cos\left(\bm{k}_{5}\cdot(\bm{r}-\bm{R}_2)\right)]^2,
\label{Vshift}
\end{eqnarray}  
where
$\bm{k}_4=k(1/3,0),
\bm{k}_5=k(0,\sqrt{3}/6),
\bm{R}_1=a(-1/4,\sqrt{3}/4),
\bm{R}_2=a(-1/4,5\sqrt{3}/4),
A=2.75,$ and $A'=1.9$.  
The first term in the potential modifies the hoppings by widening specific sites 
in the hexagon while the second term maintains a balance among hoppings at 
the opposite sides of a hexagon in the kagome lattice.  As a result of the shift  
the site specific potentials become modified to 
$V_{\text{OL}}=I_0[V_{\text{kag}}+I_sV_{\text{s}} ]+V_{z}$
(see the right panel of 
Fig.~\ref{fig1}).  The top panel in Fig.~\ref{fig2} shows that 
the resulting hoppings go from uniform, at $I_s=0$, to site-dependent, at 
finite $I_s$.  
The potential, at first glance, yields hoppings 
which approximate a tuning of $c$ according to Eqs.~\ref{conditions}.  Note 
that $t_{3,4}$ is not in accord with Eqs.~\ref{conditions}.  The 
kagome lattice in Fig.~\ref{fig1} appears to be only slightly affected but this is in 
fact not the case.  The bottom panel in Fig.~\ref{fig2} shows that 
the chemical potentials are drastically modified by $V_{\text{s}}$.  
The hopping times from site to site become prohibitively 
long with these large chemical potential shifts ($\sim E_{\text{R}}$) indicating that weak modifications 
of the hoppings ($30\%$) in this implementation of the kagome lattice lead to unwanted large shifts in the 
chemical potential.  In fact another crucial requirement in implementing 
Eq.~\ref{HamiltonianTB} is the ability to tune the chemical potential locally 
to maintain uniform renormalized chemical potentials throughout the lattice 
up to $\mathcal{O}(t^2/U_{\hexagon})$.  We find that modifying hoppings with Eq.~\ref{Vshift}  
leads to drastic and incompatible changes in chemical potentials.      

The lattice coloring scheme defined by $V_{\text{s}}$ leads to anisotropic 
interactions.  The colored lattice alters the shape of the Wannier functions in a site specific 
fashion and can therefore, in principle, induce anisotropies in the interaction.  Eqs.~\ref{conditions} 
require that bow-tie terms, $V^{\gamma}_{\alpha \beta}$,  are at least an order of magnitude 
smaller than hexagon terms, $U_{\hexagon}$.  Since these interactions cover the same 
physical distance in the kagome lattice we require anisotropies in the interaction among next 
nearest neighbors.  We find, 
by direct calculation, that band effects modify the next nearest neighbor interaction only slightly.  There is, 
at best, a $8\%$ anistropy in, e.g., the quantity $(V_{15}-V_{46})/V_{46}$, (See the 
site definitions in the top panel of Fig.~\ref{fig2}) in the range $0 \leq I_s \leq 0.4$.  
The variation in bow-tie terms compared to cross-hexagon terms was much less, $<1\%$, over the same 
range of $I_s$ values.

\section{\label{Exchange} Towards Superexchange}

We now briefly discuss practical issues in realizing low temperature 
superexchange with particles in 
optical lattices modeled by Eq.~\ref{Hamiltonian}.  Nearly all 
proposed topological lattice models make use of a delicate competition 
between anisotropic interaction terms.  Ref.~\onlinecite{Freedman0} 
invokes superexchange to generate some of these interaction terms from an 
underlying Hubbard model.  A variety of theoretical proposals in the optical lattice setting seek to 
realize the equivalent of superexchange with different techniques including the use of 
interstitials with resonant interactions \cite{Buchler}
and polar molecules \cite{Zoller_polar,Buchler_polar}.  But,  
conventionally, superexchange in a 
single band Hubbard model can be achieved with low temperatures, 
$T\ll \mathcal{O}(t^2/U$), and interaction strengths below the band 
gap, $U\ll \Delta$, where $t$ and $U$ are characteristic hopping and 
interaction energies, respectively.  The temperature sets an 
absolute energy scale which can, in principle, be experimentally tuned below the superexchange 
limit, $\mathcal{O}(t^2/U$).  In practice, however, the realization of 
sufficiently low temperatures in optical lattice setups remains elusive and has 
been a motivating factor in recent theoretical work 
(See Refs.~\onlinecite{Buchler, Zoller_polar, Buchler_polar, Trebst} for example).  We find rather 
stringent requirements on temperatures needed to 
realize conventional low temperature/single band superexchange with our direct implementation scheme, 
because, as we will show below, single band and low temperature 
requirements are mutually exclusive.  

One potential solution to the temperature problem is to simply 
increase energy scales.  The physical energy scale, 
$E_{\text{R}}$, is fixed by the mass of the constituent particles and the 
wavelength of the lasers defining the optical lattice.  
We exclude the possibility of adjusting the mass and laser wavelength.  
With more tunable experimental parameters one can, however, 
increase $t$ by lowering the lattice depth while increasing  
$U_0$ with, for example, a Feshbach resonance.  But  
note that the temperature sets a lower bound for $t$ while $\Delta$ sets an upper 
bound on interaction energies.  Typically, $t$ decreases exponentially with $\Delta$ 
for large lattice depths making the last two requirements 
in Eqs.~\ref{conditions} difficult to realize even with 
extremely low temperatures.  To quantify the parameter 
window we consider one-dimensional examples below.

\begin{figure}
\includegraphics[width=3in]{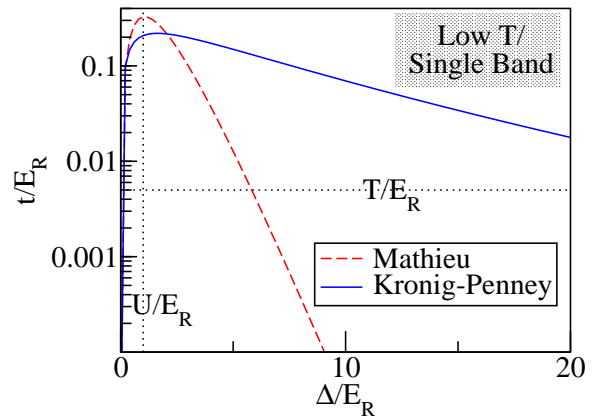}
\caption{
Hopping versus band spacing for the one-dimensional Mathieu (dashed) and Kronig-Penney (solid)
models, Eqs.~\ref{scaling}.  The vertical (horizontal) dotted line indicates 
an independently tunable interaction (temperature) scale.  The 
upper right quadrant of the graph indicates a range of hoppings and band spacings 
that yield an ideal limit for superexchange 
given the positions of the dotted lines.   
}
\label{fig3}
\end{figure}

The precise functional form relating $t$ and $\Delta$ depends on details of 
the single particle potential 
defining the lattice.  In solid state systems lattice sites are typically defined by 
spatially local atomic cores.  The resulting band gaps can be 
quite large even for large hopping strengths (band widths).  For comparison, consider a one-dimensional 
Mathieu lattice \cite{Slater} defined by interfering lasers.  Expanding the 
potential near each site yields a parabolic (harmonic) confinement 
characteristic of optical lattices.  (Note that the 
intensity profiles of standing wave laser beams will almost always yield a parabolic potential about 
each optical lattice site, as in Eq.~\ref{Vkag}) 
Consider the Kronig-Penney model \cite{KP} 
where, by contrast, individual sites are modeled by attractive delta function potentials.  For  
well separated sites ($t \ll \Delta$) the band width in the 
Kronig-Penney model scales more favorably with band spacing:
\begin{eqnarray}
t_{\text{M}}/E_{\text{R}}&\approx& \frac{4}{\sqrt{\pi}2^{3/4}} 
\left(\frac{\Delta}{E_{\text{R}}}\right)^{3/2}\exp\left(-\sqrt{2}\Delta/E_{\text{R}}\right) 
\nonumber \\
t_{\text{KP}}/E_{\text{KP}}&\approx& 
\left(\frac{\Delta}{E_{\text{KP}}}\right)\exp\left(-\frac{\pi}{2}\sqrt{\Delta/E_{\text{KP}}}\right),
\label{scaling}
\end{eqnarray}
where $E_{\text{KP}}= h/8ma_{\text{KP}}^2$ 
for the Kronig-Penney model is defined in terms of a lattice spacing $a_{KP}=a$ to draw 
an equivalence with $E_{\text{R}}$ for the optical lattice Mathieu problem.  We 
derive Eqs.~\ref{scaling} under the assumption $t \ll \Delta$.  The square 
root in the exponential suggests that Kronig-Penney-like 
systems (systems with tight confinement around 
each lattice site) are better approximated by single band models 
over a comparatively wider parameter range, provided equivalent 
energy scales.  We show this in 
Fig.~\ref{fig3} where we plot Eqs.~\ref{scaling}.  The vertical and horizontal 
dotted lines indicate an arbitrarily tunable interaction energy and temperature, respectively.  
$T/E_{\text{R}}=0.005$ and $U/E_{\text{R}}=1$ are chosen as examples.  
The upper right quadrant of the graph then corresponds to the low temperature/single 
band limit ideal for realizing superexchange.  The Kronig-Penney 
hopping and band spacing lead to a more favorable parameter window for superexchange.  Harmonically 
confined sites (e.g. Eq.~\ref{Vkag}) require much lower temperatures.  
 
\section{\label{Conclusion} Conclusion}

We have discussed potential issues
entering into our emulation scheme of a topological extended Bose-Hubbard model 
in cold atom optical lattices of dipoles finding that, 
even without considering the ring exchange terms in the 
proposed topological Hubbard model \cite{Freedman0}, it will be very 
difficult, if not impossible, to use our direct scheme to simulate the corresponding strongly correlated 
model of Ref.~\onlinecite{Freedman0}.  
We find:  
i)  The constraint that the 
superexchange energy $\mathcal{O}(t^2/U_{\hexagon})$ is much larger than the temperature
within the single-band Hubbard model (so that the band gap $\Delta$ is large 
compared with $U_0$ and $U_{\hexagon}$) 
is difficult to satisfy with currently accessible temperatures in 
experiments, $(T\sim t)$, using a direct emulation with 
harmonic optical lattices.  This is currently a problem for 
most proposals making use of superexchange in optical lattices.  A 
lattice with Kronig-Penney-like site confinement may 
allow a wider temperature window.  Other implementation 
schemes \cite{Buchler,Zoller_polar,Buchler_polar,Jan}, 
in conjunction with a kagome optical lattice \cite{Santos}, may 
also be able to avoid the prohibitively low temperature requirements. 
ii) Our suggested modifications to the kagome optical lattice,    
a tuning of hopping parameters with additional laser beams, lead to 
drastic and incompatible changes in the lattice structure itself.
By tuning the lattice to color hoppings we find large chemical potential 
shifts ($\sim E_{\text{R}}$) which correspond to prohibitively long 
hopping time scales.  
iii) We also find that, with dipoles, the 
constraints on the anisotropic interaction 
are too demanding for our direct implementation scheme.  Specifically, cross-hexagon terms 
should be an order of magnitude larger than bow-tie terms,  
$V^{\gamma}_{\alpha \beta} \ll U_{\hexagon}$, 
but end up comparable in our scheme, 
$V^{\gamma}_{\alpha \beta} \sim U_{\hexagon}$.  Tuning interaction 
anisotropy with band effects leads to only a small, $<8\%$, variation in 
next nearest neighbor interaction terms.  A low 
temperature optical lattice of polar molecules \cite{Jan} may show more 
promise in all of the above categories.  In spite of our somewhat disappointing conclusion on the
prospects of creating a topological phase with an extended Bose-Hubbard model
using kagome optical lattices, we think that it is important to continue
thinking about cold atom optical lattices as suitable systems for emulating
elusive type IIb topological phases by implementing other indirect
techniques beyond the scope of our work.  This is particularly true in view
of the highly elusive nature of type IIb topological matter, and the
fact that optical lattices allow for the possibility of emulating
Hamiltonians which are unrealistic (certainly in their pristine forms) in
solid state materials.

We thank J.I. Korsbakken, K. Park, K.B. Whaley, and C.W. Zhang 
for helpful discussions.  This work is supported by the 
Microsoft Q Project and ARO-DARPA.

\end{document}